\documentclass[journal=jctcce,manuscript=article,layout=traditional]{achemso} 

\usepackage[T1]{fontenc} 
\usepackage[version=3]{mhchem} 
\usepackage[dvipsnames]{xcolor}
\usepackage{graphicx,subcaption}
\usepackage{amssymb}
\usepackage{overpic}
\usepackage{soul}
\usepackage{comment}
\usepackage{bm}
\usepackage{ragged2e}
\usepackage{array, multirow, boldline}
\usepackage{booktabs}
\usepackage{rotating}
\usepackage{caption}

\newif\ifhighlightchanges
\highlightchangesfalse

\definecolor{darkblue}{HTML}{003D6D}

\newcommand{\bP}{{\bm P}}

\newcommand{\hbm}[1]{\hat{\bm{#1}}}

\newcommand{\bra}[1]{\left< #1\right|}
\newcommand{\ket}[1]{\left|#1 \right>}

\usepackage{accents}

\title[]{Adiabatic Electronic Momentum and the Continuity Equation with Nonlocal Effective Core Potentials in an Electronic Phase-Space Hamiltonian}

\author{Bismark Ninson}
\affiliation{Department of Chemistry, University of Rhode Island, Kingston, Rhode Island, 02881, USA}

\author{Zhen Tao}
\affiliation{Department of Chemistry, University of Rhode Island, Kingston, Rhode Island, 02881, USA}
\email{zhen.tao@uri.edu}
\begin{document}

\begin{abstract}
Adiabatic electronic momentum and current density provide fundamental information about electronic motion induced by nuclear motion under the Born-Oppenheimer separation. In practical electronic structure calculations, core electrons in heavy atoms are commonly replaced by effective core potentials (ECPs), whose nonlocal components introduce additional contributions to electronic momentum and the continuity equation.  Here, using a phase-space electronic Hamiltonian approach, we investigate these nonlocal components on adiabatic electronic momentum and the continuity equation in small molecules with moving nuclei. We demonstrate that this approach efficiently describes adiabatic electronic momentum and the divergence of electronic current density when the nonlocal contributions to the corresponding electronic operators are properly accounted for. We find that the nonlocal correction to electronic momentum is important when the ECP-centered atoms move but is small otherwise, reflecting the spatial locality of the electronic response and ECP. We further examine the implications of moving ECPs, which account for the motion of nonlocal pseudopotentials with the nuclei to restore Galilean covariance, for adiabatic electronic transport. We show that, for nondegenerate systems with time-reversal symmetry, moving ECPs do not modify the coordinate space adiabatic evolution of time-reversal even observables through second order in nuclear velocity, although they can alter the electronic momentum operator and its feedback on nuclear dynamics.
\end{abstract}

\section{Introduction}

Effective core potentials (ECPs) are routinely used in electronic structure calculations and molecular dynamics for chemical systems containing heavy elements. By replacing chemically inert core electrons with an effective potential, ECPs greatly reduce computational costs while maintaining high accuracy for valence electronic structure and properties. This concept dates back to  the pioneering work on pseudopotentials in the 1930s,\cite{Hellmann1935,Gombas1935} and has been widely developed in both  molecular quantum chemistry and solid-state electronic structure calculations.\cite{Preuss1955,Phillips1959} Conventional molecular ECPs are typically represented by angular momentum dependent semilocal potentials,\cite{Wadt1985,Andrae1990,Peterson2003,Dolg2012,Bennett2017} whereas modern plane wave methods commonly employ separable nonlocal projector representations constructed from radial projectors within each angular momentum channel.\cite{Kleinman1982,Vanderbilt1990,Goedecker1996,Hamann2013}  Despite differences in their construction and mathematical representation, both approaches introduce nonlocal electron-core interactions in coordinate space.  Throughout this work, we use the terms ECP and pseudopotential interchangeably. For a detailed survey of the historical development of pseudopotentials or ECPs, we refer readers to several excellent reviews. \cite{Pickett1989,Payne1992,schwerdtfeger_pseudopotential_2011,Dolg2012} 

Although core electrons are not treated explicitly, ECPs account for the  interactions of valence electrons with the atomic cores through an effective one-electron operator that can be decomposed into local and nonlocal components, where $\hat{V}_{\rm ECP}=\hat{V}_{\rm local}+ \hat{V}_{\rm nl}$. The consequences of this nonlocality have long been recognized in studies of electronic responses to external electromagnetic fields, where nonlocal potentials introduce additional contributions to the coupling between electrons and electromagnetic fields.\cite{starace_length_1971,baroni_ab_1986,levine_linear_1989,IsmailBeigi2001} In particular, within the velocity-gauge formulation, the first-order coupling to a spatially uniform electromagnetic vector potential involves the electronic velocity operator, which acquires an additional contribution from the commutator with the nonlocal potential,
$\dot{\hat{\bm r}}=(i/\hbar)[\hat{H}_{\rm el},\hat{\bm r}]=\hat{\bm p}/m_{\rm e}+(i/\hbar)[\hat{V}_{\rm nl},\hat{\bm r}]$. Previous studies show that nonlocal pseudopotential corrections can alter electronic velocity matrix elements and calculated optical transition strengths,\cite{Read1991,Adolph1996} as well as molecular magnetic susceptibilities and nuclear magnetic resonance (NMR) shielding constants.\cite{Pickard2003,vanWullen2012} Beyond electromagnetic perturbations, nuclear motion also induces electronic motion as the electronic wavefunction evolves with the changing nuclear geometry. Even in the adiabatic limit, this evolution gives rise to an electronic response associated with nuclear velocities. However, the importance of nonlocal ECP corrections for this nuclear motion induced electronic response remains less  investigated.

This situation is perhaps unsurprising because the adiabatic electronic response to nuclear motion cannot be calculated with the conventional Born-Oppenheimer approximation (BOA).\cite{Born1927,Born1955} Unlike explicitly propagating the electronic wavefunction in time according to the time-dependent Schr{\"o}dinger equation, BOA solves for the  instantaneous electronic eigenstates at each nuclear configuration $\bm R$. The corresponding electronic momentum and currents evaluated using these instantaneous electronic eigenstates vanish without time-reversal symmetry breaking. This problem was recognized in early studies  of vibrational optical activity\cite{Nafie1983,Stephens1985Review,Buckingham1987}. A solution has been developed using  nuclear velocity perturbation theory (NVPT)\cite{Nafie1983,Buckingham1987}, with nonadiabatic coupling $-i\hbar\bm P/M \cdot \bm d$ as the perturbation, where $\bm d$ is the derivative coupling. As the perturbation theory formally involves a sum over states, which is computationally expensive, several practical routes have been offered since then. In particular, NVPT has been implemented within density functional theory\cite{Scherrer2013NVPT} and also formulated within the exact factorization framework.\cite{Abedi2010, Scherrer2015NVPT} Further developments have extended NVPT to velocity-dependent Gaussian basis functions and plane waves, also explicitly considering nonlocal pseudopotentials.\cite{ditler_implementation_2022,ditler_vibrational_2023,kumar_calculation_2025}

An alternative framework to NVPT is to introduce the nuclear velocity dependence directly in the electronic Hamiltonian. As independently developed in the phase space electronic structure theory (PSEST)\cite{bian_phase-space_2026} and  moving Born–Oppenheimer (moving BO) approximation,\cite{barrera_moving_2026} this framework can be understood as describing the fast degrees of freedom (electrons) in a reference frame that follows the motion of the slow degrees of freedom (nuclei), yielding effective states that depend on both the coordinates and momenta of the slow degrees of freedom.  The transformation to this moving frame is governed by the adiabatic gauge potential (AGP),\cite{kolodrubetz_geometry_2017} an operator whose matrix elements are related to the derivative couplings between electronic eigenstates. For general molecular motion, however, constructing the corresponding momentum-dependent electronic Hamiltonian requires knowledge of the electronic derivative couplings, which is computationally challenging and  gauge dependent in the case of degeneracy. PSEST provides a practical approximation by replacing the non-diverging component of the derivative couplings with an effective operator $\hat{\bm\Gamma}$ constructed to satisfy the translational and rotational symmetry constraints of the exact derivative couplings.\cite{tao_basis-free_2025,bian_phase-space_2026} See Ref. \citenum{Subotnik2026PSEST} for a detailed review of the historical development of PSEST. These symmetry constraints ensure that the net momentum transfer between the electronic and nuclear degrees of freedom is captured correctly. Instead of solving for BO Hamiltonian eigenstates, the electronic phase-space Hamiltonian is diagonalized directly with the $-i\hbar\bm P/M \cdot \hat{\bm\Gamma}$ operator. Hence the wavefunction corrections due to the non-divergent component of the nonadiabatic couplings are naturally included. This approach has been shown to provide accurate all-electron results in various applications that involve electron-nuclear couplings such as adiabatic electronic momentum and current density,\cite{Tao2024CurrentDensity,tao_basis-free_2025} nuclear vibrational frequency,\cite{Bian2025Vibrational} vibrational optical activity spectra\cite{duston_phase-space_2024,Tao2024CurrentDensity,tao_non-resonant_2026} and molecular spin-rotation coupling.\cite{Peng2026SpinRotation} 

Motivated by these developments, we employ PSEST to investigate how ECPs for heavy atoms influence electronic momentum and the divergence of the current density induced by nuclear motion. In the Theory section, we first present the adiabatic evolution of electronic observables in terms of the AGP and connect this formalism to PSEST, which provides a computationally efficient approach for evaluating the corresponding electronic response. In the Results section, we report the adiabatic electronic momentum and the electronic continuity equation using PSEST for several small molecules containing heavy atoms treated with ECPs, focusing on the contributions arising from the nonlocality of the potentials. Recent work by Stengel and coworkers has demonstrated that nonlocal pseudopotentials must also account for the velocity of their associated nuclei to preserve Galilean covariance.\cite{stengel_rototranslational_2026} Motivated by this work, in the Discussion section, we examine how a moving ECP (MECP) affects adiabatic electronic properties using perturbation theory and discuss their feedback on nuclear dynamics.

\section{Theory}
\subsection{Adiabatic Evolution of Electronic Observables}
\label{sec:AGP}
We first review the adiabatic transport theory of electronic observables based on the AGP.\cite{kolodrubetz_geometry_2017} Hereafter, we use $I,J,K,L$ to label electronic states, $\alpha,\eta,\gamma,\lambda$ to denote Cartesian components ($x,y,z$), and $A,B,C,D$ to label nuclei. Bold symbols denote three-dimensional Cartesian vectors and operators. Along a nuclear trajectory $\bm R(t)$, the electronic instantaneous eigenstates $\ket{\psi_I(\bm R(t))}$ are obtained at each nuclear configuration $\bm R$, 
\begin{align}
     \hat{H}_{\rm el}(\bm R) \ket{\psi_{I}(\bm R)} &= E^{el}_{\rm I}(\bm R)\ket{\psi_{I}(\bm R)}
\end{align}
The time dependence of these electronic adiabatic eigenstates along the nuclear trajectory $\bm R(t)$ can  be viewed as their evolution through nuclear parameter space rather than as direct time evolution according to the electronic time-dependent Schr\"odinger equation. Specifically, from the chain rule, 
\begin{align}
    \frac{d}{dt}\ket{\psi_I(\bm R)} =  \sum_{A}\dot{\bm R}_A\cdot\bm \nabla_{A}\ket{\psi_I(\bm R)} \label{eq:wf1}
\end{align}
Since the instantaneous electronic eigenstates form a complete orthonormal basis, the eigenbases at different nuclear configurations $\bm R$ and $\bm R'$ are related by a unitary transformation $\hat U(\bm R,\bm R') = \sum_I\ket{\psi_I(\bm R)}\bra{\psi_I(\bm R')}$, such that $\ket{\psi_I(\bm R)}=\hat U(\bm R,\bm R')\ket{\psi_I(\bm R')}$. Holding the reference nuclear configuration $\bm R'$ fixed ($ \nabla_{A}\ket{\psi_I(\bm R')} = 0 $), changes in the instantaneous electronic eigenstates with nuclear coordinates arise entirely from changes in this unitary transformation. Hence, assuming a smoothly varying eigenbasis, the parametric derivatives of the electronic eigenstates can be represented locally by infinitesimal unitary rotations within the electronic Hilbert space. Denoting $\hat U(\bm R,\bm R')$ simply as $\hat U(\bm R)$, the adiabatic gauge potential, which generates these infinitesimal unitary rotations, is defined as
\begin{align}
\hat{\bm A}_A(\bm R)
=i\hbar\left[\bm\nabla_A\hat U(\bm R)\right]\hat U^\dagger(\bm R).
\end{align}
and the parametric instantaneous eigenstate evolution in Eq. \ref{eq:wf1} can be rewritten in terms of AGP as
\begin{align}
\frac{d}{dt}\ket{\psi_I(\bm R)}
=-\frac{i}{\hbar}\sum_A\dot{\bm R}_A\cdot
\hat{\bm A}_A(\bm R)\ket{\psi_I(\bm R)}.
\label{eq:wf2}
\end{align} 
For an electronic observable $\hat{O}$ with no explicit time dependence, the AGP therefore generates its adiabatic evolution in the nuclear parameter space
\begin{align}
   \frac{d}{dt}
    \langle \hat O\rangle
    =
    \frac{i}{\hbar}
    \sum_A
    \dot{\bm R}_A
    \cdot
    \left\langle
    \left[
    \hat{\bm A}_A,
    \hat O
    \right]
    \right\rangle
    +
    \sum_A
    \dot{\bm R}_A
    \cdot
    \left\langle
    \frac{\partial \hat O}
    {\partial \bm R_A}
    \right\rangle.
    \label{eq:AGP_transport}
\end{align}
Eq. \ref{eq:AGP_transport} resembles the Ehrenfest theorem for real-time dynamics, but here the evolution is in the nuclear parameter space, not in time, and the generator is the AGP, not electronic Hamiltonian. 

Alternatively, the same adiabatic evolution can be represented in a wavefunction response form. Inserting the resolution of the identity of the instantaneous eigenstates of $\hat H_{\rm el}(\bm R)$, $\sum_{I}\ket{\psi_I}\bra{\psi_I}$, between the operator commutator $\left[
    \hat{\bm A}_A,
    \hat O
    \right]$ in  Eq. \ref{eq:AGP_transport}, then Eq. \ref{eq:AGP_transport} can be written as
\begin{align}
   \frac{d}{dt}
    \langle \hat O\rangle
&=
    2\,\operatorname{Re}
    \left[
    \sum_A\sum_{J\neq I}
    \dot{\mathbf R}_A\cdot
    \mathbf d^{A}_{JI}\,
    O_{IJ}
    \right]+
    \sum_A
    \dot{\mathbf R}_A\cdot
    \left\langle\psi_I\left|
    \nabla_{\mathbf R_A}\hat O
    \right|\psi_I\right\rangle 
    \label{eq:wf_transport_sos1}\\
    & = 2\,\operatorname{Re}
    \left[
    \sum_A\sum_{J\neq I}
    \frac{\dot{\mathbf R}_A\cdot
    \mathbf d^{A}_{JI}}{E_I-E_J}
   [\hat H_{\rm el},\hat O]_{IJ}
    \right]+
    \sum_A
    \dot{\mathbf R}_A\cdot
    \left\langle\psi_I\left|
    \nabla_{\mathbf R_A}\hat O
    \right|\psi_I\right\rangle \label{eq:wf_transport_sos2}\\
    &  = 2\,\operatorname{Re}
    \left[\left\langle\psi_I\left|
  \frac{i}{\hbar} [\hat H_{\rm el},\hat O]\right|\delta\psi_I\right\rangle 
    \right]+
    \sum_A
    \dot{\mathbf R}_A\cdot
    \left\langle\psi_I\left|
    \nabla_{\mathbf R_A}\hat O
    \right|\psi_I\right\rangle \label{eq:wf_transport_sos3}
\end{align}
The sum excludes $J=I$ in Eq. \ref{eq:wf_transport_sos1} because the diagonal contribution to the commutator vanishes. To go from Eq. \ref{eq:wf_transport_sos1} to Eq. \ref{eq:wf_transport_sos2}, we used the relation,
\begin{align}
    \left\langle\psi_I\left|
    [\hat H_{\rm el},\hat O]
    \right|\psi_J\right\rangle = (E_I-E_J) O_{IJ} \label{eq:wf_response}
\end{align} 
Here  we define 
\begin{align}
    |\delta\psi_I\rangle  = -i\hbar\sum_{A,J\neq I}
    \frac{\dot{\mathbf R}_A\cdot
    \mathbf d^{A}_{JI}}{E_I-E_J}  |\psi_J\rangle  \label{eq:dpsi}
\end{align}
where $d^{A}_{JI} = \bra{\psi_J}\frac{\partial}{\partial \bm R_{A}}\ket{\psi_I} = -i/\hbar \bra{\psi_J}\hat{\bm A}_{A}\ket{\psi_I}$ is the derivative coupling and also  the AGP matrix elements in adiabatic basis. Eq. \ref{eq:dpsi} has the same expression as the first-order perturbed wavefunction due to nonadiabatic coupling. Eq. \ref{eq:AGP_transport} and Eq. \ref{eq:wf_transport_sos3} provide two equivalent representations of adiabatic parameter evolution. In  Eq. \ref{eq:AGP_transport}, the observable is propagated by AGP while the expectation values are evaluated in the instantaneous BO states. In Eq. \ref{eq:wf_transport_sos1}, the electronic observable is generated by the electronic Hamiltonian through $i/\hbar[\hat H_{\rm el},\hat O]$ and the expectation values are evaluated with the AGP-propagated electronic wavefunction. 

\subsection{Adiabatic Electronic Momentum and Current Density}
The theory above provides a general framework for adiabatic evolution of electronic observables in the BO framework. Among these, electronic charge density operator $\hat{\bm \rho}(\bm r) = \sum_{i}\delta(\bm r-\hat{\bm r}_i)$
and its first spatial moment $\hat{\bm r}$ are fundamental to characterizing the electronic distribution, and their adiabatic evolution provides insight into electronic flow induced by nuclear motion. To calculate their adiabatic evolution, we first set $\hat{O} = \hat{\bm r}$ and use Eq. \ref{eq:wf_transport_sos3},
\begin{align}
   m_{\rm e} \frac{d}{dt}
    \langle \hat{\bm r}\rangle = 2m_{\rm e} \,\operatorname{Re}
    \left[\left\langle\psi_I\left|
   \frac{i}{\hbar}[\hat H_{\rm el},\hat {\bm r}]\right|\delta\psi_I\right\rangle 
    \right] =   2\operatorname{Re}
    \left[\left\langle\psi_I\left|
   \hat{\bm\pi}\right|\delta\psi_I\right\rangle 
    \right]  \label{eq:drdt}
\end{align}
Here $\hat{\bm\pi} =  im_{\rm e}/\hbar[\hat H_{\rm el},\hat {\bm r}]$ is the electronic kinetic momentum, which is identical to the electronic canonical momentum $\hat{\bm p}$ in the absence of magnetic fields or nonlocal potentials. In the presence of nonlocal ECP, $\hat{\bm\pi} = \hat{\bm p} + im_{\rm e}/\hbar[\hat{V}_{\rm nl},\hat{\bm r}]$. The adiabatic electronic transport described by Eq. \ref{eq:drdt} can also be formulated in terms of Berry curvature, as established in the geometric theory of adiabatic transport.\cite{Niu1984,Resta2022,Resta2023} 

We next set $\hat{O} = \hat{\bm \rho}(\bm r)$ and  use Eq. \ref{eq:wf_transport_sos3}.
\begin{align}
    \frac{d}{dt}
    \langle \hat{\bm \rho}\rangle = 2 \,\operatorname{Re}
    \left[\left\langle\psi_I\left|
   \frac{i}{\hbar}[\hat H_{\rm el},\hat {\bm \rho}]\right|\delta\psi_I\right\rangle 
    \right] = -2\operatorname{Re}
    \left[\left\langle\psi_I\left|
   \nabla_{\bm r} \cdot \hat{ \bm j}(\bm r)\right|\delta\psi_I\right\rangle 
    \right] \label{eq:contin}
\end{align}
For a local electronic Hamiltonian, the commutator 
\begin{align}
    \frac{i}{\hbar}[\hat H_{\rm el},\hat {\bm \rho}] =\frac{i}{\hbar}[\hat{\bm p}^2/2m_{\rm e},\hat {\bm \rho}] = -\nabla_{\bm r} \cdot \hat{\bm j}(\bm r) \label{eq:contin}
\end{align} 
where $\hat{\bm j}(\bm r) = 1/{2m_{\rm e}}\sum_{i}[\delta(\bm r- \hat{\bm r}_{i})\hat{\bm p}_i+ \hat{\bm p}_i \delta(\bm r-\hat{\bm  r}_i)]$. In the presence of nonlocal ECPs, one has to take into account the additional commutator in evaluating Eq. \ref{eq:contin}
 \begin{align}
   \frac{i}{\hbar} [\hat{V}_{\rm nl},\hat {\bm \rho}] =  -\nabla_{\bm r} \cdot \hat{\bm j}_{\rm nl}(\bm r)  \label{eq:commun}
\end{align} 
The continuity equation uniquely determines the divergence of the electronic current density $\nabla_{\bm r}\cdot{\bm j}(\bm r)$, but not the full spatially resolved electronic current density ${\bm j}(\bm r) $. Unlike in the all-electron case, constructing the corresponding nonlocal current density operator $\hat{\bm j}_{\rm nl}(\bm r) $ requires an additional prescription that accounts for how electronic probability density flows between  spatial positions coupled by the nonlocal ECPs. \cite{IsmailBeigi2001,Pickard2003,Dreyer2018} We therefore focus on the continuity equation and leave the explicit construction of the nonlocal current density operator for future work.

\subsection{Phase Space Electronic Structure Theory}
As shown in the response formulation in Eq. \ref{eq:dpsi}, the adiabatic nuclear motion induces an electronic response that is linear in the nuclear velocity $\dot{\bm R}$ and also dependent on the nonadiabatic couplings. Instead of evaluating this response perturbatively, as in NVPT,\cite{Nafie1983,Scherrer2013NVPT,Scherrer2015NVPT,Luber2022} one can incorporate the AGP directly into the electronic Hamiltonian.\cite{Shenvi2009,bian_phase-space_2026,barrera_moving_2026,Subotnik2026PSEST} By doing so, the electronic Hamiltonian depends not only on the standard parameter nuclear positions, but also on how fast these parameters change, i.e. nuclear momentum $\bm P$, hence the name electronic phase space Hamiltonian. Shenvi used this Hamiltonian when conducting nonadiabatic dynamics with surface hopping.\cite{Shenvi2009}  Specifically, he constructed a superadiabatic basis by diagonalizing an electronic phase space Hamiltonian in the adiabatic basis, with nonadiabatic  couplings as the off-diagonal couplings, 
\begin{align} 
    \hat{H}_{\rm Shenvi}(\bm{R},\bP) 
    = &\sum_{IJK,A} \frac{1}{2M_A} \left( \bm{P}_A \delta_{IJ} - \bm {A}^A_{IJ} \right)\cdot
    \left( \bP_A \delta_{JK} -\bm {A}^A_{JK} \right)\ket{\psi_I}\bra{\psi_K}
     +\sum_K E_{K}(\bm{R})\ket{\psi_K}\bra{\psi_K}, 
    \label{eq:shenvi} 
\end{align}
Diagonalizing this electronic phase-space Hamiltonian  yields eigenstates $\{\phi_I\}$ that depend on both nuclear coordinates $\bm R$ and momentum $\bP$,
\begin{align}
    \hat{H}_{\rm Shenvi}(\bm{R},\bP) \ket{\phi_I(\bm{R},\bP)} &= E_{\rm I}(\bm{R},\bP)\ket{\phi_I(\bm{R},\bP)}
\end{align}    
Higher-order nonadiabatic terms can therefore be included iteratively, also known as Berry's superadiabatic construction.\cite{Berry1987,Berry1990} However, this approach is computationally expensive and is difficult to apply in cases of degeneracy, where derivative couplings between degenerate states are gauge dependent. This motivates the recent development of the electronic phase space approach,\cite{Wu2022,tao_basis-free_2025,bian_phase-space_2026} which approximates the non-divergent component of the  derivative coupling with a one-electron $\hat{\bm \Gamma}$ operator 
\begin{align}
        \hat{H}_{\rm PS}(\bm{R},\bP) 
    = & \sum_{A}\frac{1}{2M_A} \left( \bm{P}_A  - i\hbar \hat{\bm\Gamma}_A(\bm R) \right)\cdot
    \left( \bP_A - i \hbar \hat{\bm \Gamma}_A(\bm R) \right) + \hat{H}_{\rm el}(\bm R) \label{eq:Hps}
\end{align}
This $\hat{\bm \Gamma}$ operator satisfies the symmetry constraints, given below in Eqs. \ref{eq:1a}-\ref{eq:4a}, which are identical to those satisfied by the exact derivative couplings:\cite{Littlejohn2023,Wu2022} 
\begin{eqnarray}
    -i\hbar\sum_{A} {\bm d}^A_{IJ} + \bra{\psi_I}{\hat{\bm p}} \ket{\psi_J} &=& 0,\label{eq:dconstrain1a}  \\
    \sum_A \bm \nabla_{A} {\bm d}^B_{IJ}  &=& 0, 
    \label{eq:dconstrain2a}\\
    -i\hbar\sum_{A}{\bm R}_{A} \times {\bm d}^A_{IJ} + \bra{\psi_I} {\hat{\bm l}} + {\hat{\bm s}} \ket{\psi_J} &=& 0,\label{eq:dconstrain3a}\\
     -\sum_A \left({\bm R}_A \times \bm \nabla_{A} {d}^{B\eta}_{IJ}\right)_{\alpha} &
     =&  \sum_{\gamma} \epsilon_{\alpha \eta \gamma} d^{B \gamma}_{IJ},\label{eq:dconstrain4a} 
\end{eqnarray}
where $\hat{\bm l}=\hat{\bm r}\times\hat{\bm p}$ is the electronic orbital angular momentum operator and $\hat{\bm s}$ is the electronic spin operator. Eqs. \ref{eq:dconstrain1a} and  \ref{eq:dconstrain3a} encode the partition of linear and angular momentum between the nuclear and electronic degrees of freedom in the adiabatic representation and Eqs. \ref{eq:dconstrain2a} and  \ref{eq:dconstrain4a} ensure that the full adiabatic molecular Hamiltonian is translationally and rotationally invariant. 

As the exact constraints in Eqs. \ref{eq:dconstrain1a}-\ref{eq:dconstrain4a} govern only the overall momentum transfer between nuclei and electrons, they do not uniquely determine the internal electronic response, i.e., how the electrons move with individual nuclei in molecules. Here we adopt the previously developed $\hat{\bm \Gamma}$ operator as one possible prescription for partitioning the electronic response among individual nuclei, with a controllable degree of spatial locality.\cite{tao_basis-free_2025} Specifically, this $\hat{\bm \Gamma}$ operator consists of three components, $\hat{\bm \Gamma} = \hat{\bm \Gamma}' + \hat{\bm \Gamma}'' +\hat{\bm \Gamma}'''$ , each of which takes into account the coupling between electronic linear, orbital angular momentum, and spin. Take the linear momentum component for example,
\begin{align}
   \hat{\bm \Gamma}'_{A} = \frac{1}{2i\hbar}\left( \hat{\Theta}_A\hbm{p}+\hbm{p}\hat{\Theta}_A\right).\label{eq:etf}
\end{align}
Here, the locality function $\Theta_{A}(\hat{\bm r})$ provides a smooth partition of unity that assigns greater weight to nearby nuclei, with nuclear charges $Q$ as prefactors and a locality-width parameter $\sigma$. 
\begin{align}
\label{eq:theta}
    \Theta_A(\hat{\bm r}) = \frac{Q_Ae^{-|\hat{\bm r}-\bm{R}_A|^2/\sigma^2}}{\sum_B Q_Be^{-|\hat{\bm r}-\bm{R}_B|^2/\sigma^2}}.
\end{align}
Summing over nuclei $A$ in Eq. \ref{eq:etf} gives electronic momentum which satisfies Eq. \ref{eq:dconstrain1a}.

To effectively choose the degree of spatial locality $\sigma$, we further enforce a practical constraint, 
\begin{align}
    \sum_A \langle \Theta_A(\hat{\bm r}) \rangle\bm R_A - \langle \hat{\bm r}\rangle = 0 \label{eq:theta_sigma}
\end{align}
At the nonlocal limit $\sigma\rightarrow\infty$, this condition requires the center of charge to coincide with the expectation value of the electronic position operator. In the semi-local case, $ \langle\Theta_A(\hat{\bm r})\rangle \equiv Q^{\rm eff}_{A}$ can be viewed as an effective atomic charge, which was previously used to estimate shielded nuclear charges in the presence of magnetic fields. \cite{bhati_phase-space_2025} More generally, this constraint enforces the correct first moment of the electronic density when it is represented by effective charges located at the nuclear positions. Since Eq. \ref{eq:theta_sigma} depends on electronic density, this allows a different choice of $\sigma$ for different molecules. Note that since Eq. \ref{eq:theta_sigma} is a vector constraint, whereas the locality parameter $\sigma$ is a scalar, we choose $\sigma$ by minimizing the magnitude of the residual. 

 The angular momentum component $\hat{\bm \Gamma}''$ similarly partitions the electronic angular momentum to nuclear rotation, although its mathematical form is more involved. We refer readers to Ref. \citenum{tao_basis-free_2025} for the detailed expression. This construction also introduces a spatial locality parameter $\beta$ that controls the range over which neighboring nuclei contribute to the electronic angular momentum coupling. In this work, we use a large length scale of $\beta=9$ Bohr, which allows the rotational coupling to  extend throughout the small molecules studied here. This parameter choice is similar to those in previous studies. \cite{Tao2024CurrentDensity,tao_non-resonant_2026} We neglect the spin component $\hat{\bm \Gamma}'''$ because we consider only closed-shell systems, for which spin-dependent effects are beyond the scope of this work. 

\section{Results}
We solve the electronic phase space Hamiltonian in Eq. \ref{eq:Hps} using complex restricted Hartree–Fock (CRHF) theory\cite{Small2015} for a moving Kr atom and HBr, HI, ClBr, BrI and CH$_3$Br molecules, with nuclear momenta corresponding to the prescribed nuclear velocities. Using these nuclear momenta as inputs, we solve for electronic phase space eigenstates and compute the adiabatic electronic kinetic momentum $\langle\hat{\bm \pi} \rangle$ and divergence of the current density $\langle\nabla_{\bm r} \cdot \hat{\bm j}(\bm r)\rangle$ under the atomic orbital (AO) projection approximation described in Appendix A. According to Eq. \ref{eq:drdt} and Eq. \ref{eq:contin}, we then assess the accuracy of the phase space electronic structure calculations by comparing to their corresponding BO benchmarks $d\langle\hat{\bm r}\rangle/dt$ and $d\langle\hat{\bm \rho}\rangle/dt$ evaluated using finite differences. ECPs are used for systems containing heavy elements including Kr, Br and I. Specifically, we use scalar-relativistic, energy-consistent Stuttgart/Cologne ECPs in conjunction with the correlation-consistent weighted core–valence basis sets (cc-pwCVnZ-PP), where n = D, T, Q, and 5 denote double-, triple-, quadruple-, and quintuple-zeta, respectively.\cite{Peterson2003,Peterson2010} For C and H, we use the corresponding cc-pVnZ basis sets, and for Cl, we use the corresponding cc-pwCVnZ basis sets. Hereafter, we use the abbreviations nZ-PP and anZ-PP to denote these combined basis set choices.

\subsection{A Translating Kr Atom}
Before showing results for molecules, we first provide electronic momentum for exact translational motion. Here the $\hat{\bm \Gamma}$ operator is just the electronic momentum operator $ \hat{\bm p}$ according to the linear momentum constraint in Eq. \ref{eq:1a}, which is also the exact AGP as shown in Eq. \ref{eq:dconstrain1a}. This hence serves as a benchmark test to study the impact of nonlocal ECP on electronic momentum.\cite{Fatehi2011,Littlejohn2023}  

Consider a Kr atom  moving  with a nuclear velocity of $10^{-4}$ a.u, with its ionic core represented by an ECP. All the valence electrons should translate with the same velocity. In Table \ref{Tab:Kr}, we show the electronic canonical and kinetic momentum of each orbital using the cc-pwCVnZ-PP basis set series and the associated ECPs, which replace the ten core electrons occupying the first two shells ($n=1$ and $n=2$). Each occupied orbital contains two electrons and contributes $2\times10^{-4}$ a.u to the total electronic velocity. Summing over the 13 occupied orbitals gives a total of $2.6\times10^{-3}$ a.u. As shown in Table \ref{Tab:Kr}, the additional momentum correction $im_{\rm e}/\hbar[\hat{V}_{\rm nl},\hat{{\bm r}}]$ is particularly important for the $n=3$ shell but becomes less important  for electrons in shells with higher $n$. The calculated electronic momentum for electrons in 4p orbitals improves with diffuse basis functions. We also observe that the canonical momenta for degenerate orbitals with different $m_l$ are different, indicating that the nonlocal ECP velocity correction depends on $m_l$, even though the ECP used itself does not explicitly distinguish between different $m_l$ values. Together, these results show that momentum correction $im_{\rm e}/\hbar[\hat{V}_{\rm nl},\hat{{\bm r}}]$ due to the nonlocality of the ECP is important to restore the electronic kinetic momentum for a translating Kr atom.  

\begin{table*}[ht]
\centering
\caption{Electronic canonical $\langle \hat{\bm p}\rangle$ and kinetic momentum $\langle \hat{\bm\pi}\rangle$ for a translating Kr atom. All values are reported in units of $\times10^{-4}$.}
\label{Tab:Kr}
\begin{tabular}{c|cc|cc|cc|cc|cc}
\hline
& \multicolumn{2}{c|}{TZ-PP}
& \multicolumn{2}{c|}{QZ-PP}
& \multicolumn{2}{c|}{aQZ-PP}
& \multicolumn{2}{c|}{5Z-PP}
& \multicolumn{2}{c}{a5Z-PP} \\
Orb
& $\langle \hat{\bm p}\rangle$
& $\langle \hat{\bm\pi}\rangle$
& $\langle \hat{\bm p}\rangle$
& $\langle \hat{\bm\pi}\rangle$
& $\langle \hat{\bm p}\rangle$
& $\langle \hat{\bm\pi}\rangle$
& $\langle \hat{\bm p}\rangle$
& $\langle \hat{\bm\pi}\rangle$
& $\langle \hat{\bm p}\rangle$
& $\langle \hat{\bm\pi}\rangle$ \\
\hline
3s  & 3.21 & 1.94 & 3.22 & 1.92 & 3.22 & 1.92 & 3.22 & 1.92 & 3.22 & 1.95 \\[3pt]
3p  & 3.04 & 1.95 & 3.05 & 1.97 & 3.05 & 1.97 & 3.05 & 1.97 & 3.05 & 1.97 \\
  & 3.04 & 1.95 & 3.05 & 1.97 & 3.05 & 1.97 & 3.05 & 1.97 & 3.05 & 1.97 \\
 & 2.37 & 1.98 & 2.38 & 2.03 & 2.38 & 2.03 & 2.38 & 2.03 & 2.39 & 2.02 \\[3pt]
3d  & 1.92 & 1.92 & 1.99 & 1.98 & 1.99 & 1.98 & 1.99 & 1.98 & 2.01 & 2.00 \\
  & 1.92 & 1.92 & 1.99 & 1.98 & 1.99 & 1.98 & 1.99 & 1.98 & 2.01 & 2.00 \\
  & 1.56 & 1.87 & 1.67 & 1.97 & 1.67 & 1.97 & 1.67 & 1.97 & 1.69 & 2.01 \\
  & 1.56 & 1.87 & 1.67 & 1.97 & 1.67 & 1.97 & 1.67 & 1.97 & 1.69 & 2.01 \\
  & 1.44 & 1.86 & 1.56 & 1.96 & 1.56 & 1.96 & 1.56 & 1.96 & 1.59 & 2.01 \\[3pt]
4s & 2.05 & 1.94 & 2.08 & 2.02 & 2.08 & 2.01 & 2.08 & 2.02 & 2.08 & 2.01 \\[3pt]
4p & 1.81 & 1.80 & 1.97 & 1.99 & 2.02 & 2.01 & 1.97 & 1.99 & 2.02 & 2.01 \\
  & 1.81 & 1.80 & 1.97 & 1.99 & 2.02 & 2.01 & 1.97 & 1.99 & 2.02 & 2.01 \\
  & 1.68 & 1.72 & 1.89 & 1.98 & 1.97 & 2.00 & 1.89 & 1.98 & 1.97 & 2.00 \\
\hline
Sum
& 27.4 & 24.5
& 28.5 & 25.7
& 28.7 & 25.8
& 28.5 & 25.7
& 28.8 & 26.0 \\
\hline
\end{tabular}
\label{tab:basis_momentum}
\end{table*}

\begin{figure}[!htbp]
\centering\includegraphics[width=1.0\linewidth]{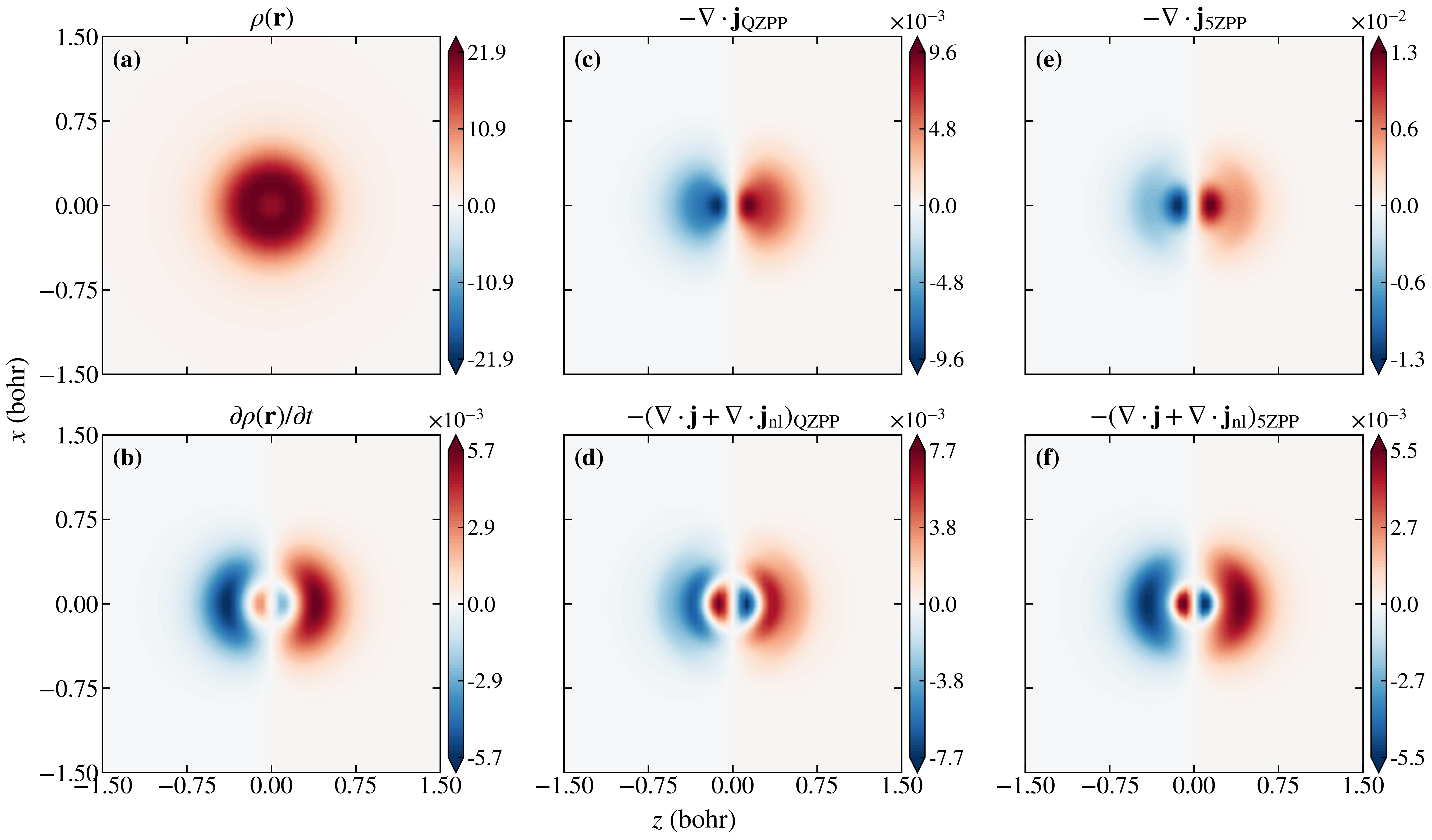}
\caption{Visualization of Kr translation along the z-axis with ECP: (a) the static valence density distribution $\rho(\bm r)$, (b) the valence density time derivative $\partial\rho(\bm r)/\partial t$, the divergence of (c) and (e)  the canonical current density $-\nabla\cdot j(\bm r)$ and (d) and (f)  the total current density $-(\nabla\cdot j(\bm r)+\nabla\cdot j_{\rm nl}(\bm r) )$ calculated with QZ-PP and 5Z-PP, respectively. All quantities are integrated over the $y$-direction and displayed in the $xz$ plane. The nonlocal component of the divergence of the current density $-\nabla\cdot j_{\rm nl}(\bm r)$  is important to match the correct nodal structures at the core region in the  density time derivative.  }
\label{fig:Kr_trans}
\end{figure}

We further test how well the continuity equation holds by comparing $d\langle\hat{\bm \rho}\rangle/dt$ and $-\nabla\cdot j(\bm r)$ plotted in Fig. \ref{fig:Kr_trans}. In Fig. \ref{fig:Kr_trans}(a) and (b), we show the electronic charge distribution $\rho(\bm r)$ and its rate of change by displacing the Kr atom with a nuclear velocity of $10^{-4}$ a.u. Because ECP replaces the core electrons, a void is seen at the center of the valence electronic charge distribution in  Fig. \ref{fig:Kr_trans}(a). Similarly, three nodes were observed along the direction of nuclear motion  in  Fig. \ref{fig:Kr_trans}(b) rather than  the  single node typically expected for the displacement of the all-electron charge density induced by nuclear motion. In Fig. \ref{fig:Kr_trans} (c)-(f), we plot the negative divergence of the current density computed using the electronic phase space approach with QZ-PP and 5Z-PP basis sets. Comparing Fig. \ref{fig:Kr_trans}(d) and (f), which include the nonlocal correction   to the divergence of current density $-\nabla\cdot j_{\rm nl}(\bm r)$, with Fig. \ref{fig:Kr_trans}(c) and (e), which drop this term, we show that $-\nabla\cdot j_{\rm nl}(\bm r)$ is essential to capture the three-nodal structure in the valence electronic charge distribution in  Fig. \ref{fig:Kr_trans}(b). 

\subsection{Molecular Vibrations}
We also computed electronic momentum for a series of small molecules including diatomics such as HBr, BrCl, HI and BrI and a polyatomic molecule CH$_3$Br. For all the molecules studied here, we moved the nuclei according to a  thermal kinetic energy at $T=298$K. For diatomic molecules, including  HBr, BrCl, and HI,  we moved the heavy nuclei along and perpendicular to the bonding axis, labeled as "on-axis" and "off-axis", respectively. For BrI and CH$_3$Br, we moved the nuclei along their respective vibrational modes, labeled as $\nu_1$ and $\nu_1$-$\nu_9$, respectively.  We determine the locality parameter $\sigma$ for each molecule using Eq. \ref{eq:theta_sigma}, which correspond to 1.2, 1.2, 1.5, 1.8, and 1.2 Bohr for HBr, HI, ClBr, BrI, and  CH$_3$Br respectively. The nuclear charges $Q$ in Eq. \ref{eq:theta} are 1, 
17, 35, and 53 a.u. for H,  Cl, Br, and I, respectively. 

\begin{figure}[!htbp]
\centering\includegraphics[width=1.0\linewidth]{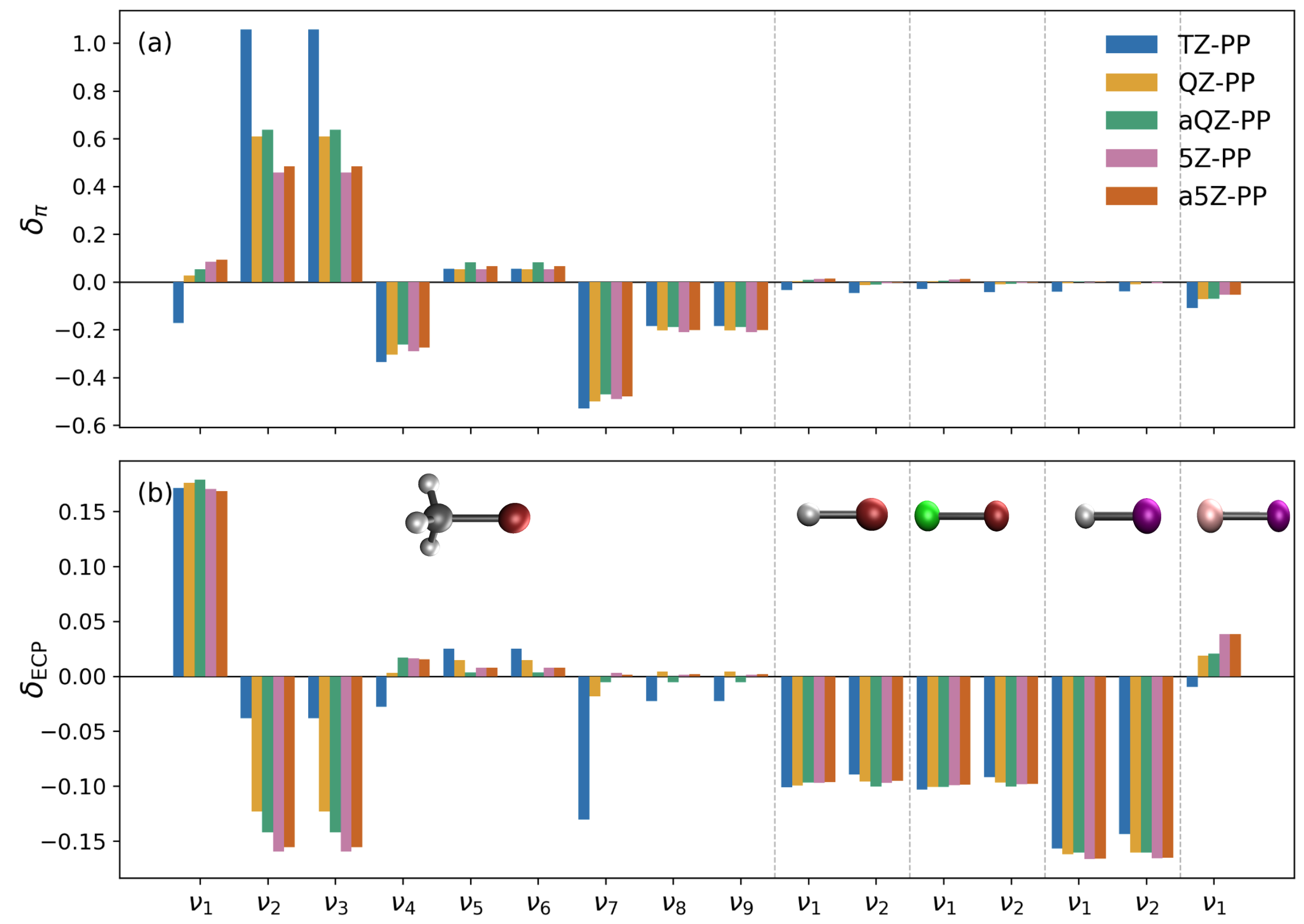}
\caption{Comparison of (a) electronic kinetic momentum $\langle\hat{\bm \pi}\rangle$ computed using the phase space approach with the finite-difference electronic velocity. The signed relative error $\delta_\pi =
{\left\|\langle\hat{\bm \pi}\rangle\right\|}
/{\left\|m_{\rm e}{d\langle\hat{\bm r}\rangle}/{dt}\right\|}-1$ is reported. Comparison of  (b) the ECP correction $im_{\rm e}/\hbar\langle[\hat{V}_{\rm nl},\hat{{\bm r}}]\rangle$ to the canonical momentum $\langle\hat{{\bm p}}\rangle$. The ratios $\delta_{\rm ECP} = \pm \left\|im_{\rm e}/\hbar\langle[\hat{V}_{\rm nl},\hat{{\bm r}}]\rangle\right\|/\left\|\langle\hat{{\bm p}}\rangle\right\|$ are reported with the signs representing parallel (positive) or anti-parallel (negative) to the direction of the canonical electronic momentum. The x-axis labels correspond to molecular motions, including vibrational modes $\nu_1$-$\nu_9$ of CH$_3$Br, on-axis $\nu_1$  and off-axis $\nu_2$ moving Br for HBr, BrCl, on-axis $\nu_1$  and off-axis $\nu_2$ moving I for HI, and the stretching vibrational mode $\nu_1$ of BrI, from left to right. The electronic basis sets used here are TZ-PP (blue), QZ-PP (yellow), aQZ-PP (green), 5Z-PP (magenta), and a5Z-PP (orange) with ECPs used for heavy atoms Br and I. }
\label{fig:mol}
\end{figure}

\begin{figure}[!htbp]
\centering\includegraphics[width=1.0\linewidth]{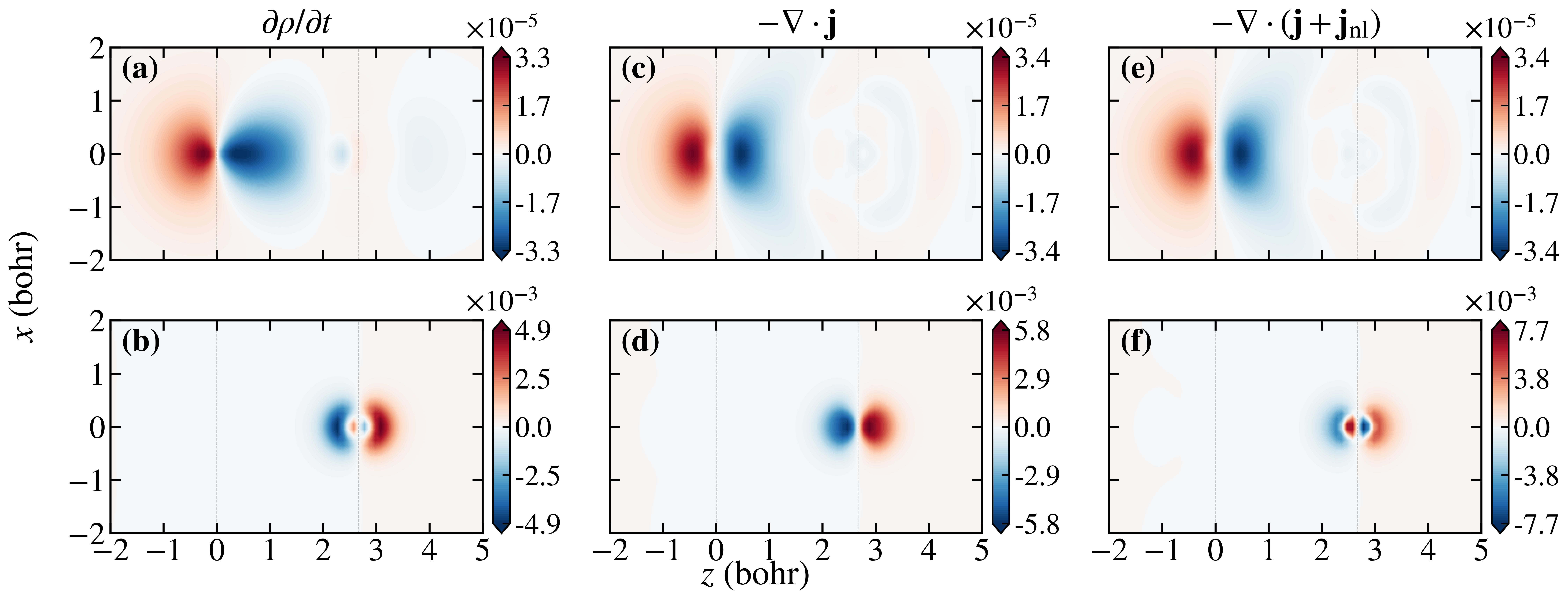}
\caption{Continuity equation visualization for (a)(c)(e) moving H and (b)(d)(f) moving Br in the HBr molecule. The H atom is centered at the origin and the Br atom is centered at 2.669 Bohr, as indicated by the dashed gray lines. The BO benchmarks $d\rho(\bm r)/dt$ in (a)-(b) are obtained using finite differences. The negative divergence of the canonical electronic current density, $-\nabla \cdot j(\bm r)$ computed with phase space is shown in (c)-(d). The nonlocal ECP corrections are further included  in (e)-(f). These corrections are essential for reproducing the correct nodal structure when Br moves, as shown in (f), but are negligible when only the H atom moves, as shown in (e). Here cc-pwCVQZ-PP and cc-pVQZ basis sets are used for Br and H, respectively.} 
\label{fig:HBr}
\end{figure}

We show the computed adiabatic electronic momentum associated with the nuclear motion specified above in Fig. \ref{fig:mol}.  In Fig. \ref{fig:mol}(a), we plot the signed relative errors in the computed electronic kinetic momentum with respect to benchmark $\delta_\pi =
{\left\|\langle\hat{\bm \pi}\rangle\right\|}
/{\left\|m_{\rm e}\dfrac{d\langle\hat{\bm r}\rangle}{dt}\right\|}-1$. According to Fig. \ref{fig:mol}(a), better agreement is seen with higher angular momentum basis sets, and the additional diffuse basis functions have only a small effect. To investigate the importance of the nonlocal correction to the electronic momentum for these nuclear motions, we plot the ratio between the correction and the canonical momentum $\delta_{\rm ECP} = \pm \left\|im_{\rm e}/\hbar\langle[\hat{V}_{\rm nl},\hat{{\bm r}}]\rangle\right\|/\left\|\langle\hat{{\bm p}}\rangle\right\|$ in Fig. \ref{fig:mol}(b), where the signs indicate if the ECP correction is parallel (positive) or anti-parallel (negative) to the canonical electronic momentum. We have verified that the two contributions are nearly collinear for all nuclear motions considered. Here we see that the correction is mostly around $10\%$-$15\%$ of the regular momentum for the nuclear motions that involve the heavy nuclei motion. For higher vibrational modes $\nu_{n}$ of CH$_3$Br, where $n>3$, the magnitude of the ECP correction is much smaller compared to the canonical momentum. This observation can be understood from the spatial locality of the electronic response: the adiabatic electronic motion is predominantly associated with the moving nuclei, whereas the nonlocal ECP correction is concentrated near the Br ECP center and is particularly sensitive to its motion.  As shown in Table 2 in Appendix B, the Br nuclear velocity is significantly smaller than those of the other nuclei for higher vibrational modes.  Hence, the relatively small velocity of Br in these vibrational modes leads to a weak ECP correction compared to the canonical electronic momentum. This spatial locality is further illustrated in Fig. \ref{fig:HBr}, where we examine the continuity equation for HBr with H moving (panels (a), (c), and (e)) and Br moving (panels (b), (d), and (f)), separately. The nonlocal ECP correction is negligible when H moves but becomes essential when Br moves, restoring the correct nodal structure of the electronic density response and agreement with the finite-difference BO benchmark.

\section{Discussion: Moving Effective Core Potentials}
The electronic Hamiltonians used to compute the electronic momentum and currents above are based on static ECPs.  More recently, Stengel and co-workers have taken into account of moving nuclei and showed that the conventional nonlocal pseudopotentials fail to satisfy Galilean covariance when the nuclei move.\cite{stengel_rototranslational_2026} To restore Galilean covariance, they showed that the nonlocal pseudopotentials must move with the nuclei,
i.e., for a nucleus $A$, $\hat{V}^{A}_{\rm nl} \rightarrow e^{im_{\rm e}/\hbar\dot{\bm R}_{A}\cdot{(\bm r-\bm r')}}\hat{V}^{A}_{\rm nl}(\bm r,\bm r')$. When written in electronic operator form, this can be written as
\begin{align}
    \hat{V}^{A}_{\rm MECP} \equiv e^{im_{\rm e}/\hbar\dot{\bm R}_{A}\cdot \hat{\bm r}}\hat{V}^{A}_{\rm nl}e^{-im_{\rm e}/\hbar\dot{\bm R}_{A}\cdot \hat{\bm r}} =  \hat{V}^{A}_{\rm nl}+\frac{im_{\rm e}}{\hbar}\dot{\bm R}_{A} \cdot [\hat{\bm r},\hat{V}^{A}_{\rm nl} ]  + O(\dot{\bm R}_{A}^2)
\end{align} 
In the first section, we review the effect of the moving ECP on electronic adiabatic properties and show that its contribution to $d\langle\hat{\bm r}\rangle/dt$ enters beyond the second order in nuclear velocity perturbation for constant nuclear velocity and hence can be neglected through second order. Then we discuss the electronic feedback on nuclear dynamics, and show that the effect of moving ECP cannot be neglected in dynamics. 
\subsection{Electronic Adiabatic Propagation}
\label{sec:EAP}
We consider the leading-order effect of the moving ECP as a perturbation to the static ECP electronic Hamiltonian.
\begin{align}
\hat H(\bm R, \dot{\bm R}) &= \hat H_{\rm el}(\bm R)
+\hat H_{\rm MECP}^{(1)}(\bm R, \dot{\bm R})+ O(\dot{\bm R}^2) \\
    \hat H_{\rm MECP}^{(1)} (\bm R, \dot{\bm R})&= \frac{im_{\rm e}}{\hbar}\sum_{A}\dot{\bm R}_{A} \cdot [\hat{\bm r},\hat{V}^{A}_{\rm nl} ] \label{eq:HMECP1}
\end{align}
Here, $\hat H_{\rm el}$ is the BO electronic Hamiltonian containing the static ECPs. Treating $\hat H_{\rm MECP}^{(1)}$ as a perturbation, we expand the instantaneous electronic eigenstates and their corresponding energies in powers of nuclear velocity,
\begin{align}
|\psi_I\rangle &= |\psi_I^{(0)}\rangle + |\psi_I^{(1)}\rangle + O(\dot{\bm R}^2) ,\\
E_I &= E_I^{(0)} + E_I^{(1)} + O(\dot{\bm R}^2) .
\end{align}
Here, $|\psi_I^{(0)}\rangle$ and $E_I^{(0)}$ are the eigenstate and eigenvalue of the static-ECP Hamiltonian, respectively. For nondegenerate electronic states, standard first-order perturbation theory gives
\begin{align}
|\psi_I^{(1)}\rangle &= \sum_{J\ne I}|\psi_J^{(0)}\rangle
\frac{\langle\psi_J^{(0)}|\hat H_{\rm MECP}^{(1)}|\psi_I^{(0)}\rangle}{E_I^{(0)}-E_J^{(0)}},\\
E_I^{(1)} &= \langle\psi_I^{(0)}|\hat H_{\rm MECP}^{(1)}|\psi_I^{(0)}\rangle.
\end{align}
As the zeroth-order eigenfunctions are real and $\hat H_{\rm MECP}^{(1)}$ is purely imaginary in this representation, its diagonal matrix elements vanish, yielding $E_I^{(1)}=0$. Furthermore, the first-order corrections to the electronic eigenfunctions $|\psi_I^{(1)}\rangle$ are purely imaginary.

We next examine how the moving ECP perturbation modifies the adiabatic electronic response. Because the moving ECP introduces nuclear momentum dependence into the electronic Hamiltonian, the adiabatic response generally involves both nuclear coordinate space and nuclear momentum space AGPs. Here, we focus on the nuclear coordinate space contribution under the assumption of constant nuclear velocity, leaving the effect of nuclear momentum space AGP for future studies.  According to Eq.~\ref{eq:dpsi}, the electronic wavefunction response is already linear in nuclear velocity. Therefore, the first order MECP corrections to the electronic eigenstates and couplings contribute at second order in nuclear velocity. We therefore expand
\begin{align}
|\delta\psi_I\rangle &= |\delta\psi_I^{[0]}\rangle + |\delta\psi_I^{[1]}\rangle + O(\dot{\bm R}^3) ,
\end{align}
where $|\delta\psi_I^{[0]}\rangle=O(\dot {\bm R})$ is the conventional static ECP response in Eq. \ref{eq:dpsi} and $|\delta\psi_I^{[1]}\rangle=O(\dot{\bm R}^2) $ is the leading MECP correction.

Expanding the electronic states, energy differences, and derivative couplings to first order in the MECP perturbation, we obtain
\begin{align}
|\delta\psi_I^{[0]}\rangle &= -i\hbar\sum_{A,J\ne I}\frac{\dot{\bm R}_A\cdot\bm d_{JI}^{A,(0)}}{\Delta E_{IJ}^{(0)}}|\psi_J^{(0)}\rangle,\\
|\delta\psi_I^{[1]}\rangle &=  -i\hbar\sum_{A,J\ne I}\dot{\bm R}_A\cdot\left[
\frac{\bm d_{JI}^{A,(1)}}{\Delta E_{IJ}^{(0)}}|\psi_J^{(0)}\rangle
+\frac{\bm d_{JI}^{A,(0)}}{\Delta E_{IJ}^{(0)}}|\psi_J^{(1)}\rangle
-\frac{\bm d_{JI}^{A,(0)}\Delta E_{IJ}^{(1)}}{(\Delta E_{IJ}^{(0)})^2}|\psi_J^{(0)}\rangle
\right],
\end{align}
where $\Delta E_{IJ}^{(n)}=E_I^{(n)}-E_J^{(n)}$, and
\begin{align}
\bm d_{JI}^{A,(0)} &= \langle\psi_J^{(0)}|\nabla_{\bm R_A}\psi_I^{(0)}\rangle,\\
\bm d_{JI}^{A,(1)} &= \langle\psi_J^{(1)}|\nabla_{\bm R_A}\psi_I^{(0)}\rangle
+\langle\psi_J^{(0)}|\nabla_{\bm R_A}\psi_I^{(1)}\rangle.
\end{align}
 Since $E_I^{(1)}=0$, the first-order correction to the energy difference vanishes, $\Delta E_{IJ}^{(1)}=0$. Moreover, $\bm d_{JI}^{A,(0)}$ is real whereas $\bm d_{JI}^{A,(1)}$ is purely imaginary. Consequently, $|\delta\psi_I^{[0]}\rangle$ is purely imaginary  and $|\delta\psi_I^{[1]}\rangle$ is real under the stated assumptions.

Substituting these expansions into the commutator representation of the nuclear coordinate space adiabatic response (similar to Eq. \ref{eq:wf_transport_sos3}) gives
\begin{align}
\left.
\frac{d}{dt}
\langle\hat{\bm r}\rangle
\right|_{\dot{\bm R}}
=
2\operatorname{Re}\
\Big[
&
\langle\psi_I^{(0)}|
\frac{i}{\hbar}[\hat H_{\rm el},\hat{\bm r}]
|\delta\psi_I^{[0]}\rangle
+
\langle\psi_I^{(1)}|
\frac{i}{\hbar}[\hat H_{\rm el},\hat{\bm r}]
|\delta\psi_I^{[0]}\rangle
+
\langle\psi_I^{(0)}|
\frac{i}{\hbar}[\hat H_{\rm MECP}^{(1)},\hat{\bm r}]
|\delta\psi_I^{[0]}\rangle
\nonumber\\
&+
\langle\psi_I^{(0)}|
\frac{i}{\hbar}[\hat H_{\rm el},\hat{\bm r}]
|\delta\psi_I^{[1]}\rangle
\Big]
+O(\dot{\bm R}^3).\label{eq:drdt_R}
\end{align}
The first term is the conventional adiabatic response of the static ECP Hamiltonian. The remaining three terms arise specifically from the moving ECP and represent the correction to the instantaneous electronic eigenstate, the explicit correction to the electronic Hamiltonian, and the correction to the adiabatically propagated wavefunction, respectively. Under time reversal, $i/\hbar[\hat H_{\rm el},\hat{\bm r}]$, $|\delta\psi_I^{[0]}\rangle$, and $|\psi_I^{(1)}\rangle$ change sign, while $i/\hbar[\hat H_{\rm MECP}^{(1)},\hat{\bm r}]$ and $|\delta\psi_I^{[1]}\rangle$ remain unchanged. Hence, all three nominally second-order matrix elements are imaginary and vanish after taking the real component in Eq. \ref{eq:drdt_R}.

The coordinate-space adiabatic response reduces to
\begin{align}
\left.
\frac{d}{dt}
\langle\hat{\bm r}\rangle
\right|_{\dot{\bm R}}
&=
2\operatorname{Re}\,
\langle\psi_I^{(0)}|
\frac{i}{\hbar}[\hat H_{\rm el},\hat{\bm r}]
|\delta\psi_I^{[0]}\rangle
+O(\dot{\bm R}^3). \label{eq:drdt_R2}
\end{align}
Thus, the leading order coordinate space adiabatic response is determined by the static ECP Hamiltonian and its AGP generated wavefunction response, which already incorporate the effects of static ECP nonlocality. Under the stated time-reversal symmetry and nondegeneracy assumptions, the additional MECP contributions vanish through second order in nuclear velocity. Note that Eq. \ref{eq:drdt_R} is general for any operator $\hat{O}$, but Eq. \ref{eq:drdt_R2} is only true for  time-reversal even operators such as $\hat{\bm r}$ or $\hat{\bm \rho}$ studied here. For a time-reversal odd operator  $\hat{O}$, the operator $i/\hbar[\hat H_{\rm el},\hat{ O}]$ is time-reversal even. Hence, the first order response in nuclear velocity  vanishes but the second order terms from the moving ECP can contribute.

\subsection{Electronic Feedback on Nuclear Dynamics}
In the previous section, we showed that the second-order MECP correction to coordinate space adiabatic electronic velocity vanishes for nondegenerate systems with a time-reversal symmetric BO Hamiltonian. Nevertheless, moving ECPs can play an important role in the electronic feedback on nuclear dynamics.

Within the conventional adiabatic approximation, the
electronic feedback can manifest as a
geometric contribution through the diagonal component of the adiabatic gauge potential, also known as the Berry connection,
$\bm A_{II}=i\hbar \bm d_{II}$ for a single electronic eigenstate $I$. The resulting  effective nuclear Hamiltonian has the form,\cite{Mead_Truhlar_1979,mead:1992:rmp}
\begin{align}
    \hat{H}_{\rm mol}
    =
    \sum_A
    \frac{
    \left(\bm P_A-\bm A_{II}^{A}\right)^2
    }{2M_A}
    + E_{I}(\bm R),
\end{align}
The corresponding nuclear equations of motion are
\begin{align}
\dot{\bm R}_A
&=
\frac{\bm P_A-\bm A_{II}^{A}}{M_A},
\\
M_A\ddot{\bm R}_A
&=
-\nabla_{\bm R_A}E_{I}
+
\bm F_A^{\rm Berry},
\end{align}
where the Berry force is determined by the curvature
of the Berry connection,
\begin{align}
F_{A\alpha}^{\rm Berry}
&=
\sum_{B\eta}
\left(
\frac{\partial A^{B\eta}_{II}}
{\partial R_{A\alpha}}
-
\frac{\partial A^{A\alpha}_{II}}
{\partial R_{B\eta}}
\right)
\dot R_{B\eta}.
\end{align}
The total linear and angular kinetic momenta are conserved\cite{Bian2023}
\begin{align}
    \bm P_{\rm tot} &= \langle \hat{\bm p}\rangle + \sum_{A}M_A\dot{\bm R}_A  =  \sum_{A}\bm P_A  \\
    \bm L_{\rm tot} &= \langle \hat{\bm l}\rangle + \langle \hat{\bm s}\rangle+\sum_{A}M_A\bm R_A \times \dot{\bm R}_A  =  \sum_{A}\bm R_A \times \bm P_A 
\end{align}
The final equalities follow from the phase conditions in Eq. \ref{eq:dconstrain1a} and Eq. \ref{eq:dconstrain3a} applied to the diagonal matrix elements of electronic eigenstate $I$, where 
 \begin{align}
  \sum_{A}A^{A\alpha}_{II} &=  \langle \hat{ p}_{\alpha}\rangle_{II} \label{eq:A1}\\
  \sum_{A}\Big(\bm R_A \times \bm A^{A}_{II}\Big)_{\alpha} &=  \langle \hat{ l}_{\alpha}\rangle_{II}+\langle \hat{ s}_{\alpha}\rangle_{II}   \label{eq:A2} 
 \end{align}
The total linear and angular momenta are hence represented by the nuclear canonical linear and angular momentum, which are conserved due to the translational and rotational invariance of the potential energy surface.\cite{Bian2023} For a nondegenerate spinless BO Hamiltonian with
time-reversal symmetry, the electronic eigenstates
can be chosen real, so the conventional Berry
connection and Berry curvature vanish. The nuclear
equations of motion therefore reduce to conventional
BO dynamics.

However, as demonstrated by Stengel et al.,\cite{stengel_rototranslational_2026}
moving nonlocal pseudopotentials introduce an additional
contribution to the effective Berry connection,
\begin{align}
\bm A_{II}^{A,\rm eff}
&=
\bm A_{II}^{A,\rm BO}
+
\bm A_{II}^{A,\rm nl}.
\end{align}
This additional gauge connection $\bm A_{II}^{A,\rm nl} = \bra{\psi_I}im_{\rm e}/\hbar[\hat{V}^{A}_{\rm nl},\hat{\bm r}]\ket{\psi_I}$ can generate a
nonzero Berry curvature and hence modify nuclear dynamics. They further derived the total conserved momentum from the Ehrenfest Lagrangian formalism\cite{stengel_rototranslational_2026}
\begin{align}
    \bm P^{\rm MECP}_{\rm tot} &= \langle \hat{\bm \pi}\rangle + \sum_{A}M_A\dot{\bm R}_A  \label{eq:Ptot}\\
    \bm L^{\rm MECP}_{\rm tot} &= \langle \hat{\bm l}^{\rm MECP}\rangle + \langle \hat{\bm s}\rangle + \sum_{A}M_A\bm R_A \times \dot{\bm R}_A \label{eq:Ltot}
\end{align}
where $ \hat{\bm l}^{\rm MECP} =\hat{\bm r} \times \hat{\bm p} + im_{\rm e}/\hbar\sum_{A}\bm R_{A}  \times [\hat{V}^{A}_{\rm nl},\hat{\bm r}] $. As adding the moving pseudopotentials in the electronic Hamiltonian does not modify the phase conditions in Eqs. \ref{eq:dconstrain1a}-\ref{eq:dconstrain4a}, Eqs. \ref{eq:A1}-\ref{eq:A2} still hold and hence the electronic canonical momentum similarly cancels the corresponding BO Berry connection components in the total momentum expressions in Eqs. \ref{eq:Ptot}-\ref{eq:Ltot}. For the total linear momentum, the additional electronic momentum correction $im_{\rm e}/\hbar[\hat{V}_{\rm nl},\hat{\bm r}]$ further cancels  the additional Berry gauge connection $\sum_{A}\bm A^{A,\rm nl} $. An analogous cancellation occurs for the total angular momentum.  Hence the total linear $\bm P^{\rm MECP}_{\rm tot}$ and angular momentum  $ \bm L^{\rm MECP}_{\rm tot}$ can be similarly represented by the nuclear canonical linear $\bm P^{\rm MECP}_{\rm tot}=\sum_A \bm P_{A} $ and angular momentum $\bm L^{\rm MECP}_{\rm tot}=\sum_{A}\bm R_A \times \bm P_A $.

\subsection{Electronic Phase Space Hamiltonian and Moving Pseudopotentials}
If we add the moving pseudopotential correction from Stengel et al.\cite{stengel_rototranslational_2026} to the electronic phase space Hamiltonian in Eq. \ref{eq:Hps}, the nuclear kinetic momentum becomes
\begin{align}
    \dot{\bm R}_A
&=
\frac{\bm P_A-\bm A_{II}^{A,\rm nl}-i\hbar \bm \Gamma^{A}_{II}}{M_A},
\end{align}
where the BO connection is approximated by $i\hbar \bm \Gamma^{A}_{II}$. In this work, we use the previously developed $\hat{\bm \Gamma}$ that satisfies the exact symmetry constraints in Eqs. \ref{eq:dconstrain1a}-\ref{eq:dconstrain4a},\cite{tao_basis-free_2025} which are also given here for clarity, 
 \begin{eqnarray}
    -i\hbar\sum_{A} \hat{\bm \Gamma}_{A} + {\hat{\bm p}}  &=& 0,\label{eq:1a}  \\
    \left[
    -i\hbar\sum_A\bm\nabla_A+\hat{\bm p},
    \hat{\bm\Gamma}_{B}
    \right]
    &=& 0, 
    \label{eq:2a}\\
   -i\hbar\sum_A
    \left(\bm R_A\times\hat{\bm\Gamma}_{A}\right)
    +\hat{\bm l}+\hat{\bm s}
    &=& 0,\label{eq:3a}\\
     \left[
    -i\hbar\sum_A
    \left(\bm R_A\times\bm\nabla_A\right)_\alpha
    +\hat l_\alpha+\hat s_\alpha,
    \hat\Gamma_{B\eta}
    \right]
    &=&
    i\hbar\sum_\gamma
    \epsilon_{\alpha\eta\gamma}
    \hat\Gamma_{B\gamma}.\label{eq:4a} 
\end{eqnarray}
The total conserved linear and angular momenta are consistent with the conserved linear and angular momentum in Eqs. \ref{eq:Ptot} - \ref{eq:Ltot}.

During the preparation of this manuscript, Ohad et al.\cite{ohad_momentum_2026} independently developed a different phase space electronic structure formulation incorporating nonlocal pseudopotentials. They treat the moving ECP correction and BO adiabatic transport in the moving nuclei on the same level by adding nonlocal pseudopotentials through a nonlocal $\hat{\bm \Gamma}$ correction,  
\begin{align}
    \hat{\bm \Gamma}^{\rm nl}_{A} = \frac{m_e}{\hbar^2}[\hat{V}^{A}_{\rm nl},\hat{\bm r}]\label{eq:Gnl}
\end{align}
 in the phase space electronic Hamiltonian in Eq. \ref{eq:Hps}. The corresponding eigenstates hence include both the moving pseudopotentials and the beyond-BO correction.\cite{ohad_momentum_2026} To distinguish their formulation and notation, we refer to their approach as a moving pseudopotential electronic phase space (MPEPS) approach.  With the addition of $\hat{\bm \Gamma}^{\rm nl}$, the symmetry constraints of $\hat{\bm \Gamma}$ are modified,
 \begin{eqnarray}
    -i\hbar\sum_{A} \Big(\hat{\bm \Gamma}_{A} + \hat{\bm \Gamma}^{\rm nl}_{A}\Big)+ {\hat{\bm \pi}}  &=& 0,\label{eq:1}  \\
   -i\hbar\sum_A
    \left(\bm R_A\times\hat{\bm\Gamma}_{A}\right)
    +\hat{\bm l}_{\rm MPEPS}+\hat{\bm s}
    &=& 0,\label{eq:3}
\end{eqnarray}
 where 
  \begin{align}
     \hat{\bm l}_{\rm MPEPS} =\hat{\bm r} \times \hat{\bm \pi} =  \hat{\bm r} \times \hat{\bm p} + \hat{\bm r} \times im_{\rm e}/\hbar\sum_{A} [\hat{V}^{A}_{\rm nl},\hat{\bm r}]\label{eq:LMPEPS}
 \end{align} and the two other constraints that ensure the translational and rotational invariance of the MPEPS energy follow similarly.
 The resulting conserved total linear momentum has the same form as our expression in Eq. \ref{eq:Ptot}, which is consistent with Stengel et al.\cite{stengel_rototranslational_2026} The conserved total angular momentum, however,  differs from Eq. \ref{eq:Ltot}, where their electronic angular momentum is defined in terms of the kinetic momentum operator as given in Eq. \ref{eq:LMPEPS}.
By treating the MECP correction and the conventional AGP as two perturbations on equal footing, Ohad et al. derive a modified Nafie relation for the electronic kinetic momentum evaluated in  the MPEPS eigenstates,\cite{ohad_momentum_2026}
\begin{align}
    \langle \hat{\bm \pi}\rangle_{\rm MPEPS} = m_{\rm e} \frac{d}{dt}
    \langle \hat{\bm r}\rangle_{\rm BO} + m_{\rm e}\Big \langle \Big[\hat{\bm r},\sum_{A}\frac{\bm P_{A}}{M_A}\cdot \hat{\bm \Gamma}^{\rm nl}_{A}\Big]\Big\rangle_{\rm BO} \label{eq:pi_MPEPS}
\end{align}

While Ohad et al. aim to develop a phase space electronic structure theory incorporating moving ECPs, our focus here is on understanding how moving ECPs modify conventional BO adiabatic electronic propagation, which guides our choice of $\hat{\bm \Gamma}$ in this work.  Starting  from the conventional BO  AGP formalism for electronic adiabatic transport and examining the effects of MECP correction  perturbatively under the assumption of constant nuclear velocity, we find that this correction vanishes through second order in nuclear velocity for the adiabatic evolution of time-reversal even observables, but can play a key role for time-reversal odd observables. (Sec. \ref{sec:EAP}) It would be interesting to further explore the implications of these different perspectives, particularly how they can lead to different partitions of angular momentum between electronic and nuclear degrees of freedom.

\section{Conclusion and Future Directions}
In this work, we have investigated adiabatic electronic momentum and the electronic continuity equation in molecular systems with ECPs, using an electronic phase space Hamiltonian approach. The nonlocal components of ECPs introduce an additional contribution to the electronic momentum operator, $im_{\rm e}/\hbar\langle[\hat{V}_{\rm nl},\hat{{\bm r}}]\rangle$, which formally needs to be included to describe electronic motion consistently. Our numerical results show that the correction is important for nuclear motion involving ECP-centered atoms but can be negligible otherwise, due to the spatial locality of both the adiabatic electronic response and the nonlocal ECPs. By employing a simple prescription for selecting a single spatial locality parameter for each molecule, we obtain adiabatic electronic momentum in good agreement with benchmarks based on the rate of change of the electronic position expectation value. We further demonstrate consistency with the continuity equation by evaluating the divergence of electronic current density. In the discussion section, we examine the impact of the recently published moving ECP\cite{stengel_rototranslational_2026} perturbatively on the adiabatic electronic transport within the nuclear coordinate space and find the correction vanishes through second order in nuclear velocity for the adiabatic evolution of time-reversal even observables, but can play an important role for time-reversal odd observables. These corrections   can further play an important role in total momentum conservation and electronic feedback on nuclear dynamics. We further distinguish our present work with a recently released, independent development of electronic phase space approach incorporating moving ECP by Ohad et al.\cite{ohad_momentum_2026}

Several directions remain for future investigation. First, extending the present analysis to spatially resolved electronic current density would provide a more complete picture of electronic motion in the presence of nonlocal ECPs, beyond the divergence of the current density considered here. Second, a more detailed investigation of adiabatic electronic response in nuclear velocity space would clarify the role of moving ECPs and their associated gauge potentials. Third,  incorporating moving ECPs into electronic phase space dynamics would enable a more comprehensive assessment of their effects on coupled electron-nuclear motion and momentum conservation. Finally, although the present study focuses on small closed-shell molecules, a particularly interesting direction is to examine adiabatic spin transport in systems with large spin-orbit couplings, where explicitly incorporating the spin-dependent contributions of relativistic ECPs becomes important.\cite{Dolg2012} By consistently treating momentum transfer between electronic and nuclear degrees of freedom, the electronic phase space approach presents an attractive framework for investigating spin-vibrational coupling, spin-phonon interactions and potentially chiral induced spin selectivity.\cite{bloom_ciss_2024}

\section*{Acknowledgments}
We thank Joseph Subotnik and Xuezhi Bian for helpful discussions on moving pseudopotentials, and Anatoli Polkovnikov, Anushya Chandran, and Bernardo Barrera for insightful discussions on adiabatic gauge potentials. This work was supported in part by the Rhode Island Foundation Medical Research Grant. 

\section*{Appendix}
\appendix
\section{Finite Atomic-Orbital Basis Approximation to the Nonlocal ECP Commutator}
The nonlocal ECP contribution to the electronic kinetic momentum is evaluated in a finite atomic orbital (AO) basis ${\chi}$. Here the Greek indices $\mu$, $\nu$, and $\kappa$ label AOs. We define the projector onto the AO space as
\begin{align}
\hat P_{\mathrm{AO}}
= \sum_{\mu\nu}
|\chi_\mu\rangle
(S^{-1})_{\mu\nu}
\langle\chi_\nu|,
\end{align}
where $\{\chi_\mu\}$ are the AO basis functions and
$S_{\mu\nu}=\langle\chi_\mu|\chi_\nu\rangle$
is the AO overlap matrix.
The matrix elements of the nonlocal ECP commutator are
approximated by inserting this projector between the operators,
\begin{align}
&\langle\chi_\mu|
[\hat V_{\mathrm{nl}},\hat r_\alpha]
|\chi_\nu\rangle\approx
\langle\chi_\mu|
\hat V_{\mathrm{nl}}\hat P_{\mathrm{AO}}\hat r_\alpha
|\chi_\nu\rangle-
\langle\chi_\mu|
\hat r_\alpha\hat P_{\mathrm{AO}}\hat V_{\mathrm{nl}}
|\chi_\nu\rangle.
\end{align}
Similarly, the divergence of
the nonlocal electronic current density $-\nabla\cdot\mathbf j_{\mathrm{nl}}(\bm r)$ in Eq. \ref{eq:commun} is evaluated in real
space using the finite AO-basis projection, 
\begin{align}
-(\nabla\cdot\mathbf j_{\mathrm{nl}})_{\mu\nu}
(\mathbf r)
&=\frac{i}{\hbar}
\langle\chi_\mu|
[\hat V_{\mathrm{nl}},\hat\rho(\mathbf r)]
|\chi_\nu\rangle\\
& \approx\frac{i}{\hbar}
\Big(
\langle\chi_\mu|
\hat V_{\mathrm{nl}}\hat P_{\mathrm{AO}}
\hat\rho(\mathbf r)
|\chi_\nu\rangle-
\langle\chi_\mu|
\hat\rho(\mathbf r)\hat P_{\mathrm{AO}}
\hat V_{\mathrm{nl}}
|\chi_\nu\rangle
\Big)\\
&=\frac{i}{\hbar}
\sum_{\kappa}\Big[
(\mathbf V_{\mathrm{nl}}\mathbf S^{-1}
)_{\mu\kappa} \chi_\kappa((\mathbf r)
\chi_\nu(\mathbf r)-
\chi_\mu(\mathbf r)\chi_{\kappa}(\mathbf r)
(\mathbf S^{-1}\mathbf V_{\mathrm{nl}})_{\kappa\nu}
\Big].
\end{align}
where $\chi_\mu(\mathbf r)$ denotes an AO basis function
evaluated at position $\mathbf r$.
This approximation neglects contributions from intermediate
states outside the finite AO space, described by the
complementary projector
$\hat Q_{\mathrm{AO}}=1-\hat P_{\mathrm{AO}}$,
and becomes exact in the complete basis limit. Despite the approximation introduced here, our data in Table \ref{Tab:Kr} and Fig. \ref{fig:mol} show reasonable convergence with larger basis sets.

\section{Nuclear Velocities for CH$_3$Br Vibrational Modes}
\begin{center}
\captionof{table}{Cartesian components of nuclear velocities for the
vibrational modes of CH$_3$Br.}
\label{tab:nuclear_velocities}
\scriptsize
\resizebox{\textwidth}{!}{%
\begin{tabular}{ccrrrrrrrrr}
\toprule
Atom & Component
& $v_1$ & $v_2$ & $v_3$ & $v_4$ & $v_5$
& $v_6$ & $v_7$ & $v_8$ & $v_9$ \\
\midrule
\multirow{3}{*}{C}
& $x$ & -0.159830 & -0.062059 & -0.043393 &  0.066361 &  0.041917 & -0.029978 &  0.025613 & -0.057047 & -0.040147 \\
& $y$ &  0.187269 & -0.050610 & -0.040225 & -0.077754 &  0.034149 & -0.027739 & -0.030010 & -0.046510 & -0.037197 \\
& $z$ & -0.007464 &  0.059093 & -0.080022 &  0.003099 & -0.040791 & -0.054028 &  0.001196 &  0.054660 & -0.073552 \\
\midrule
\multirow{3}{*}{H}
& $x$ & -0.113953 &  0.511425 &  0.214919 & -0.456203 &  0.320868 &  0.071969 &  0.295913 &  0.374636 &  0.114318 \\
& $y$ &  0.156041 & -0.465044 & -0.088646 &  0.281025 & -0.074099 &  0.202684 &  0.450650 &  0.588511 &  0.181659 \\
& $z$ & -0.010914 & -0.081778 &  0.201532 &  0.041635 &  0.176414 &  0.662719 & -0.184151 & -0.232667 & -0.088687 \\
\midrule
\multirow{3}{*}{H}
& $x$ & -0.127695 & -0.211077 &  0.426690 & -0.301544 & -0.481284 &  0.291674 & -0.190555 &  0.186877 & -0.186925 \\
& $y$ &  0.145120 &  0.508610 & -0.268631 &  0.403935 & -0.141686 &  0.499173 &  0.064045 & -0.076594 &  0.056696 \\
& $z$ &  0.009347 & -0.074193 &  0.127292 & -0.186380 & -0.197571 &  0.005886 &  0.533054 & -0.531429 &  0.483120 \\
\midrule
\multirow{3}{*}{H}
& $x$ & -0.133878 &  0.027866 & -0.412112 & -0.231969 & -0.385966 &  0.027117 & -0.409400 &  0.113310 &  0.547518 \\
& $y$ &  0.138834 &  0.224100 &  0.570017 &  0.474670 & -0.229345 & -0.340286 & -0.158453 &  0.038258 &  0.201648 \\
& $z$ & -0.015970 & -0.156557 &  0.094396 &  0.098526 &  0.552862 &  0.035639 & -0.363101 &  0.117517 &  0.475623 \\
\midrule
\multirow{3}{*}{Br}
& $x$ &  0.029099 &  0.005245 &  0.003667 &  0.002549 &  0.000604 & -0.000432 & -0.000012 &  0.000057 &  0.000040 \\
& $y$ & -0.034094 &  0.004277 &  0.003400 & -0.002986 &  0.000492 & -0.000400 &  0.000014 &  0.000046 &  0.000037 \\
& $z$ &  0.001359 & -0.004994 &  0.006763 &  0.000119 & -0.000588 & -0.000778 & -0.000001 & -0.000054 &  0.000073 \\
\bottomrule
\end{tabular}%
}
\end{center}

\bibliography{Ref}

\end{document}